\documentclass[twocolumn,nobibnotes,pra]{revtex4}
\usepackage{graphicx}
\usepackage{amsmath}
\usepackage{color}
\usepackage{doi}
\begin{document}

\newcommand{\rval}{4.2(5)}

\title{Polarization of electric field noise near metallic surfaces}
\date{\today}

\author{Philipp Schindler}
\email{pschindler@berkeley.edu}
\author{Dylan J Gorman}
\author{Nikos Daniilidis}
\author{Hartmut~H\"affner}

\affiliation{\mbox{Dept. of Physics, University of California, Berkeley, CA 94720, USA}}

\begin{abstract}
  Electric field noise in proximity to metallic surfaces is a poorly
  understood phenomenon that appears in different areas of
  physics. Trapped ion quantum information processors are particular
  susceptible to this noise, leading to motional decoherence which
  ultimately limits the fidelity of quantum operations. On the other
  hand they present an ideal tool to study this effect, opening new
  possibilities in surface science. In this work we analyze and
  measure the polarization of the noise field in a micro-fabricated
  ion trap for various noise sources. We find that technical noise
  sources and noise emanating directly from the surface give rise to
  different degrees of polarization which allows us to differentiate
  between the two noise sources. Based on this, we demonstrate a
  method to infer the magnitude of surface noise in the presence of
  technical noise.
 \end{abstract}

\maketitle

A future scalable ion-trap quantum computer seems to require ion
shuttling operations and thus micrometer scale
ion-traps~\cite{Kielpinski2002}. It has become evident that trapping
an ion close to a conducting surface gives rise to unexpected large
electric field noise resulting in decoherence of the motional modes of
the ion and thus limiting the quality of many-body quantum
operations~\cite{Turchette2000,Monroe1995a,Hite2012,Daniilidis2011,Allcock2011a,Deslauriers2006a}.
A comprehensive review on this topic is given in
Ref.~\cite{Brownnutt2014}.

While there are strong experimental hints that this anomalous noise is
due to surface effects, the exact mechanisms remain
unknown~\cite{Brownnutt2014,Safavi-Naini2011,Dubessy2009}. Moreover,
many of the predictions of micrscopic models describing the noise have
yet to see experimental verification. One prediction, common to a
large class of noise models, is that the noise perpendicular to the
trap surface is larger than parallel to it. In this work, we
investigate this polarization effect, and use it to disentangle
surface noise from other sources.

Ion traps designed for quantum information processing usually provide
three dimensional confinement due to a time-varying field along two
axes and an electrostatic field along the
third~\cite{Wineland1998}. The effect of the time varying electric
field can be approximated by a pseudopotential, resulting in a
three-dimensional harmonic potential.  Any electric field noise at the
ion position creates a fluctuating force ultimately destroying the
coherence of the ion motion. In a quantum computing setting, many-body
quantum operations are enabled by the motion of the ions where the
collective modes act as a quantum bus~\cite{Cirac1995a}. Thus the
achievable gate fidelity is limited by electric field noise, which can
be observed and quantified as an increased heating rate of the ion
motion~\cite{Wineland1998,Brownnutt2014}.

The heating rate for the motion of a single mode $k$ is determined by
the projection of the electric field fluctuations onto the direction
of the respective normal mode given by the unit vector
$\hat{e}_k$. More specifically:
\[ \dot{n} =  \frac{e^2}{4 m_I \hbar \omega_t} \vec{S}_E \cdot \hat{e}_k(\omega)
\]
with ion mass $m_I$, trap frequency $\omega_t$ and the power spectral
density of the electric field fluctuations along the mode direction
$\vec{S}_E \cdot
\hat{e}_k$~\cite{Brownnutt2014,Wineland1998,Turchette2000}.
The ion is thus sensitive to a particular direction of the noise which
can be used to measure the noise polarization.
To our knowledge, no systematic measurements of the noise polarization
have been performed. Moreover, we show that such polarization
measurements give us the possibility to distinguish technical noise
sources from surface noise. Thus, an experimenter is able to decide
whether improving the electronics will reduce the heating rate or
whether the heating rate originates from excessive surface noise on
the trap. This is a crucial information as upgrading electronics is
easier than altering the surface of the trap.

The method is especially useful for experiments aiming at uncovering
the source of surface noise which might suffer from unknown
contributions of technical noise.  Such experiments investigate the
effect of various surface
treatments~\cite{Hite2012,Daniilidis2014}. If no effect is observed
after surface treatment, it might not be clear whether the treatment
was ineffective, or whether the noise was dominated by technical
sources masking the effect from the surface.
With this in mind, we seek to establish an method to distinguish
surface effects from technical noise.  The quantity of interest will
be the degree and the direction of the polarization. A large class of
technical noise will be strongly polarized under a particular angle
given by the electrode geometry. In contrast, surface noise is
expected to be relatively
unpolarized~\cite{Low2011,Brownnutt2014,Dubessy2009}. 
% Thus, measuring
% the degree and the direction of the polarization will allow us to
% exclude certain technical noise models.

This manuscript is structured as follows: In section~I we investigate
the expected polarization from surface noise. In section~II we show
how to distinguish surface noise from technical noise originating from
voltage sources. Section~III describes the measurement method to infer
the noise polarization and finally in section~IV we present a method
to extract the surface noise in the presence of technical noise.

\section{Polarization of surface noise}

In the following we will describe the expected polarization from
microscopic surface noise models. For this we will evaluate the noise
spectral density 
\[\vec{S}_E \sim \mathcal{F} \bigl{(} \langle \vec{E}(0) \vec{E}(\tau) \rangle \bigr{)}
\]
which is proportional to the Fourier transform of the autocorrelation
function of the electric field. For an ion above a large surface, the
polarization of the noise is given by the ratio of the noise spectral
density perpendicular and parallel to the trap $R =
\vec{S}_E\cdot\hat{e}_z / \vec{S}_E\cdot\hat{e}_x$.

One possible cause for surface noise are fluctuating patch
potentials~\cite{Dubessy2009}. For such patch potentials it has been
shown that in the limits of both infinitely large and small patches the
polarization cannot exceed $R=2$~\cite{Low2011}. The derivation from
reference~\cite{Dubessy2009} considers arbitrary ion-surface distances
but neglects the polarization of the noise as the absolute value of
the noise spectral density is used. It is straightforward to extend
this treatment to a directional noise spectral density~\footnote{See
  equation (6), (9) and (A1) in reference~\cite{Dubessy2009}}:
 \begin{multline}
   \vec{S}_E(\omega, d) \sim  \\
   \int_0^\infty \int_0^{2\pi} dk  d\theta k^3 e^{-2dk} S_\xi(k \cos \theta, k \sin \theta)   \begin{pmatrix}\cos \theta \\ \sin \theta \\ 1\end{pmatrix} ^2
   \label{eq:1}
 \end{multline}
 where $S_\xi(k \cos \theta, k \sin \theta)$ is the two dimensional
 Fourier transform of the spatial correlation function of the
 patches. Assuming an exponential autocorrelation function with
 correlation length $\xi$ leads to \[ S_\xi(k \cos \theta, k \sin
 \theta) = \frac{2 \pi \xi^2}{(1+\xi^2 k^2)^{3/2}} \, . \] Thus the
 integral over $\theta$ in equation~\ref{eq:1} can be evaluated
 resulting in a polarization of $R=2$, independent of the ion-surface
 distance and the correlation length of the patches.

 Another possible noise model is based on fluctuating dipoles on the
 surface. Possible sources for these dipoles are
 two-level-fluctuators~\cite{Martinis2005}, fluctuating adatomic
 dipoles~\cite{Safavi-Naini2011,Safavi-Naini2013}, or
 adatom-diffusion~\cite{Wineland1998}. The expected noise polarization
 is independent of the microscopic origin of the dipoles provided that
 the mean distance between two dipoles is much smaller than the
 ion-surface distance. For dipoles without spatial correlations, the
 autocorrelation function of the electric field can be expanded around
 the ion position $(0,0,d)$ in a second order approximation leading to
\begin{multline*}
  \langle \vec{E}(0) \vec{E}(\tau) \rangle \sim 
\langle p(0) p(\tau) \rangle \int dx dy \,
 \frac{1}{r_0^{5}} \begin{pmatrix} d \cdot x \\ d \cdot  y \\ x^2 + y^2 -    2 d^2 \end{pmatrix}^2  \\
 \sim  \frac{1}{d^4} \langle p(0) p(\tau) \rangle \begin{pmatrix} 1 \\ 1 \\
    2\end{pmatrix}
\end{multline*}

with $ \langle p(0) p(\tau) \rangle $ being the autocorrelation of the
dipole magnitude and $r_0 = \sqrt{x^2+y^2+z^2}$. The spectral noise
density is proportional to the Fourier transform of the field
autocorrelation leading to
\[\vec{S}_E(\omega,d) \sim  \mathcal{F} \bigl{(} \langle \vec{p}(0) \vec{p}(\tau) \rangle \bigr{)} \frac{1}{d^4} \begin{pmatrix} 1 \\ 1 \\
    2\end{pmatrix} \; .
\]
Therefore we also find a noise polarization $R=2$.

\section{Modeling technical noise}

Another important class of noise sources is technical noise. In the
following we will denote technical noise as all noise that is not
directly related to effects on the trap surface. Notorious sources for
technical noise are fluctuating voltage supplies driving the
electrodes. Often, suitable sources are not commercially available and
therefore several experimental groups have developed custom voltage
sources, which are compared to batteries as a low-noise
reference~\cite{Bowler2013,Baig2013,Harlander2010,Poschinger2009}.

However, employing a low-noise voltage source does not warrant the
absence of technical noise. Examples of such noise sources are Johnson
noise from the low-pass filters on the trap electrodes, or pickup of
electric fields on the wiring to the trap electrodes. While Johnson
noise can be reduced by careful engineering of the filter electronics,
it is often difficult to estimate the magnitude of the noise caused by
pick-up of environmental fields and the magnitude of the induced field
might vary on a daily timescale.  For example, we observed that
electromagnetic shielding via a Faraday cage around an ion trap
apparatus reduced the noise by one order of
magnitude~\cite{Daniilidis2014}.  
% In the following, we will present a
% method to distinguish between technical noise models and surface noise
% by measurements on the ion itself. For this, we analyze two models for
% technical noise and their influence on the polarization of the noise
% at the ion position.

Technical noise can be modeled by applying fluctuating voltages to
each electrode. We consider two different noise models: {\it(i)} a
voltage independent model where the magnitude of the noise on all
electrodes is equal and {\it(ii)} a voltage dependent model where the
noise magnitude is proportional to the applied voltage. Model {\it(i)}
describes for instance Johnson noise originating from the filter
electronics. In contrast model {\it(ii)} is suitable to describe
fluctuating voltage references of the individual digital to analog
converters.  For both models, we assume that the noise on different
electrodes is temporally uncorrelated, i.e. there are no fixed phase
relation between the corresponding voltages as for instance expected
if the noise would be caused by electronic pickup.

\begin{figure}
  \includegraphics[width = \columnwidth ]{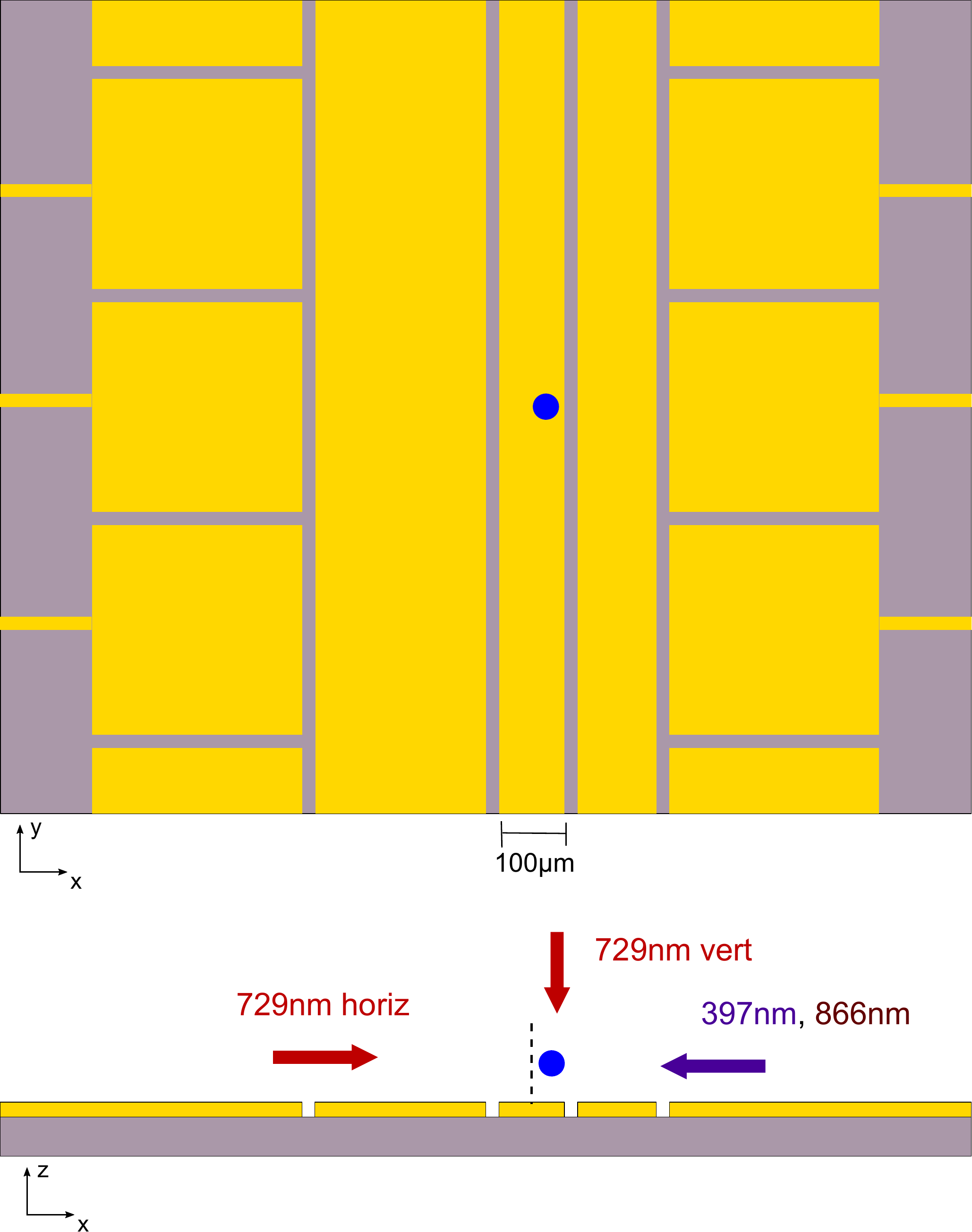}
  \caption{ (color online) Schematic drawing of the asymmetric surface
    trap used in the experiments. The orientation of the cooling
    lasers is illustrated where the beams at wavelengths 397nm and
    866nm perform Doppler cooling and the laser at 729nm enables
    sideband cooling close to the motional ground state. The expected
    trapping height is 107$\mu$m above the surface.}
  \label{fig:trap}
\end{figure}

The contribution of each electrode to the heating of the ion motion
can be determined by evaluating the electric field that a certain
voltage on the electrode generates at the ion position. Since the
noise on the electrodes is assumed to be temporally uncorrelated and
the wavelength is much larger than the ion-surface distance, the total
noise at the ion position is proportional to the sum of the squares of
the electric fields of all electrodes, projected on the respective
normal mode direction. For planar ion-trap geometries as shown in
figure~\ref{fig:trap}, the contribution from the central electrode,
directly below the ion, dominates over all other electrodes. This
effect can be exploited to distinguish technical noise from surface
noise in such a planar ion trap.

This effect is especially striking for the voltage independent noise
model {\it(i)} in an asymmetric trap where the two RF rails have
considerably different widths as sketched in
figure~\ref{fig:trap}. This geometry leads to a trapping position
which is not centered on the central electrode. Thus, the electric
field originating from the central electrode at the trapping position
does not point perpendicular to the trap surface but rather at an
angle $\phi_g \approx 15^\circ$. Since the noise is dominated by the
central electrode, the noise is maximal if the mode axis is
approximately aligned with $\phi_g$. The noise contribution of the
central electrode is about a factor of 60 larger than that of the
electrode with the second largest contribution.  For the voltage
dependent noise model {\it(ii)} the angle of the maximum noise depends
on the applied static voltages and needs to be analyzed for each
particular set of voltages.

\begin{figure}
  \includegraphics[width = \columnwidth ]{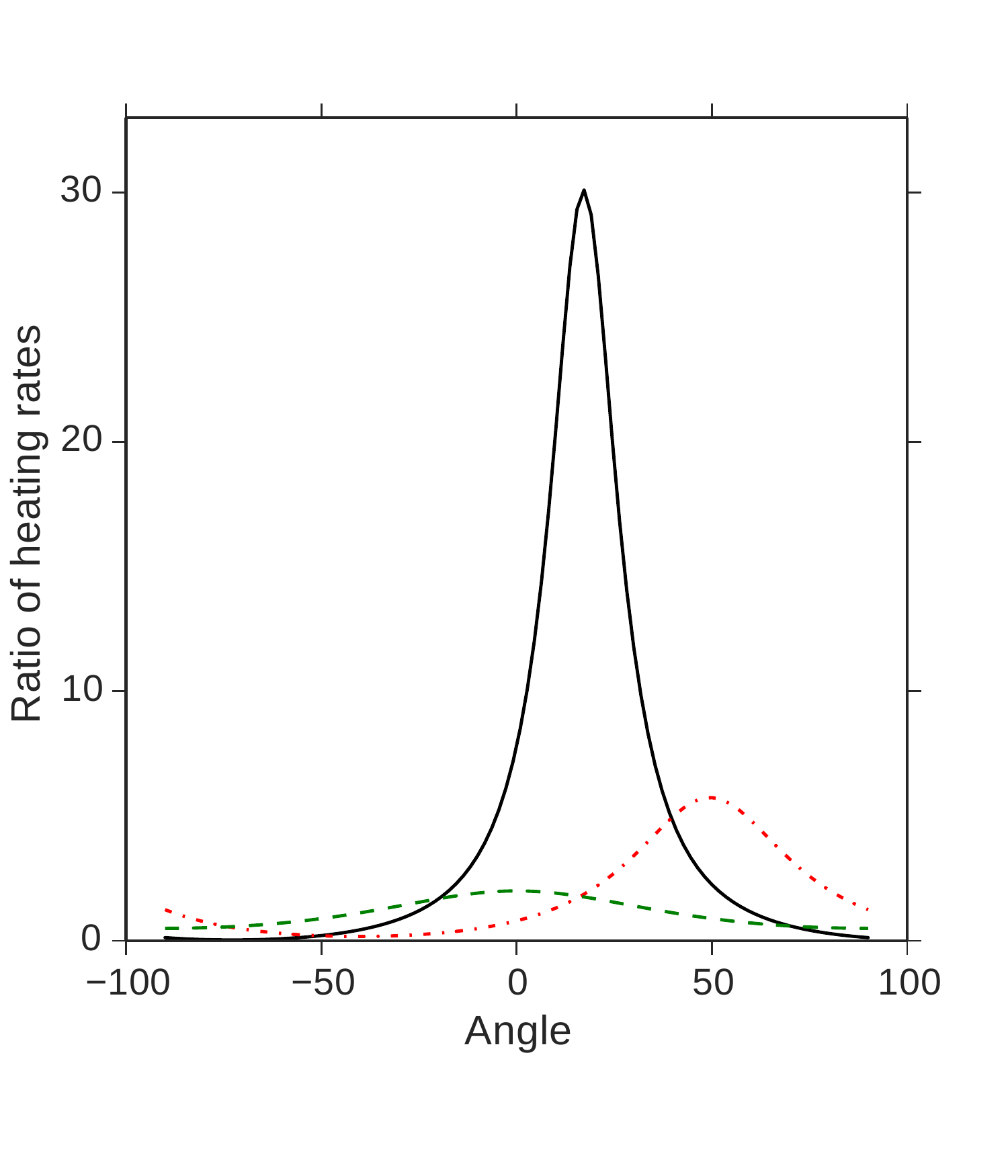}
  \caption{(color online) Ratio of the predicted heating rates in the two radial modes
    for various noise models. The green dashed line corresponds to
    surface noise, the black solid line corresponds to the voltage
    independent noise model and the red dashdotted line corresponds
    to the voltage dependent model.}
  \label{fig:ratio_th}
\end{figure}

The noise polarization can be determined independently of the absolute
noise magnitude by evaluating the ratio of the heating rates in two
normal modes while rotating them. The black solid line in
Fig.~\ref{fig:ratio_th} shows the expected ratio of the heating rate
of the two radial modes for the voltage independent noise model,
leading to a maximum ratio of $R_{\rm indep}\approx 30.1$, which can
be observed at an angle of $\phi_{\rm indep} \approx 17^\circ$.  For
the voltage dependent noise level and the set of voltages used in our
setup, the maximum ratio is $R_{\rm dep}\approx 5.7$ for an angle
$\phi_{\rm dep} \approx 50^\circ$.

Another possible noise source is electromagnetic pickup, which is
expected to be largely common to all electrodes but the surrounding
ground. Thus, the trap can be treated as a single electrode, resulting
in an electric field pointing almost perpendicular to the trap surface
with the maximum angle being close to 4$^\circ$ and a maximum ratio of
$R_\textrm{pickup} > 10^5$. The angle deviating from zero is due to
the asymmetric geometry of the electrodes.

\section{Measuring noise polarization}

The polarization of the noise in an ion-trap can be estimated by
measuring the heating rates of the normal modes while rotating the
mode direction with respect to the trap surface. In order to reduce
systematic errors in the measurements, it is advisable to perform all
measurements at approximately the same mode frequency. Thus, it is
beneficial to use the two radial modes, as their frequencies are close
whereas the axial trap frequency is usually considerably smaller than
any radial frequency. We denote the two normal radial modes
$\updownarrow$ and $\leftrightarrow$, where the mode $\leftrightarrow$
shows an angle $\phi$ with respect to the trap surface. The heating
rates for those two modes are given by
\begin{align*} S_{\updownarrow} = S_{\rm max} \cos(\phi - \phi_{\rm max})^2 + S_{\rm min} \sin(\phi-\phi_{\rm max})^2 \\
  S_{\leftrightarrow} = S_{\rm max} \sin(\phi - \phi_{\rm max})^2 +
  S_{\rm min} \cos(\phi-\phi_{\rm max})^2 \;, \end{align*} where
$S_{\rm max, min}$ is the maximum (minimum) noise amplitude and
$\phi_{\rm max}$ is the angle where the maximum noise can be
observed. 

The radial trap modes can be rotated without affecting their
frequencies by altering the static confinement. For this, the
potential of each electrode $k$ is expanded in first and second order
spherical harmonic expansion $U_k(\vec{r}) \approx \sum_i C_{i,k}
Y_i(\vec{r})$, where $C_{i,k}$ are the expansion coefficients and
$Y_i(\vec{r})$ are the spherical harmonic
functions~\cite{Littich2011,Allcock2010}. We then find a set of
voltages $V_{i,k}$ by finding solutions to the linear equation
$\vec{e}_i = \sum_k C_{i,k} V_{i,k}$ with $\vec{e}_i$ being the basis
vector corresponding to applying the spherical harmonic $Y_i(\vec{r})$.
It is sufficient to consider only the spherical harmonic functions
$Y_0(\vec{r}) = z^2 - x^2 - y^2$, $Y_1(\vec{r}) = x^2-y^2$ and
$Y_2(\vec{r}) = x \cdot y$. The functions $Y_{1,2}(\vec{r})$ allow control
over the orientation of the radial trap axes as they correspond to a
trapping configuration where the radial trap axes are aligned with the
trap surface, $\phi_1=0^\circ$, or at $\phi_2=45^\circ$
respectively. 

The orientation of the radial modes can be aligned to an arbitrary
angle~$\phi$ by applying a set of voltages that generates a potential
with coefficients
\begin{align*}
C_1 = C \, \cos(2\phi) \\
C_2 = C \, \sin(2\phi)
\end{align*}
where $C$ has to be large enough to overcome symmetry breaking due to
stray fields. The resulting potential including the confinement in the
axial direction is then: \[ U = C_0 (x^2+y^2-2z^2) + C_1 (x^2-y^2) +
C_2 (x \cdot y) + U_\textrm{RF} \] with $C_0$ determining the strength
of the axial confinement. The RF potential $U_\textrm{RF}$ and $Y_0$
have rotational symmetry around $z$ and thus do not affect the mode
orientation.

In an asymmetric surface trap, such as shown in figure~\ref{fig:trap},
it is also possible to rotate the radial trap axes by applying a
static negative bias voltage onto the RF drive. This orients the trap
axis of the higher frequency mode ($\updownarrow$) to $\phi_g$, which
corresponds to the orientation where one of the normal modes is
aligned with the field from the central electrode and hence close to
the orientation of the maximum noise for voltage independent noise
yielding an expected ratio of $R \approx 29.3$.

In order to estimate the ratio $R$, the heating rates of both radial
modes need to be measured. The heating rates of a each mode can be
inferred by comparing the relative strength of the blue and the red
sideband after cooling the mode near the motional ground
state~\cite{Leibfried2003}. Detecting these sidebands requires a
geometry where the laser wave-vector has a considerable overlap with
the respective mode axis. In order to provide this, we use the
laser-beam geometry as sketched in figure~\ref{fig:trap}. Having a
laser beam perpendicular to the trap surface is not advisable when
using a laser which causes charging on the trap
surface~\cite{Allcock2011a,Harlander2010}. Of particular concern is
the laser light at 397~nm, required to prepare both modes at a
reasonably low temperature to allow for sideband-cooling. To
circumvent shining this light directly onto the trap, we apply a
parametric coupling of the two modes \cite{Gorman2014}.

The laser required to perform ground-state cooling and analysis of the
final state has a wavelength of 729~nm which does not charge the trap
surface notably~\cite{Harlander2010}. We also verified that the laser
light does not affect the heating rate. We therefore can perform
ground-state cooling and temperature analysis with a vertical 729~nm
beam.

\begin{figure}
  \includegraphics[width = \columnwidth ]{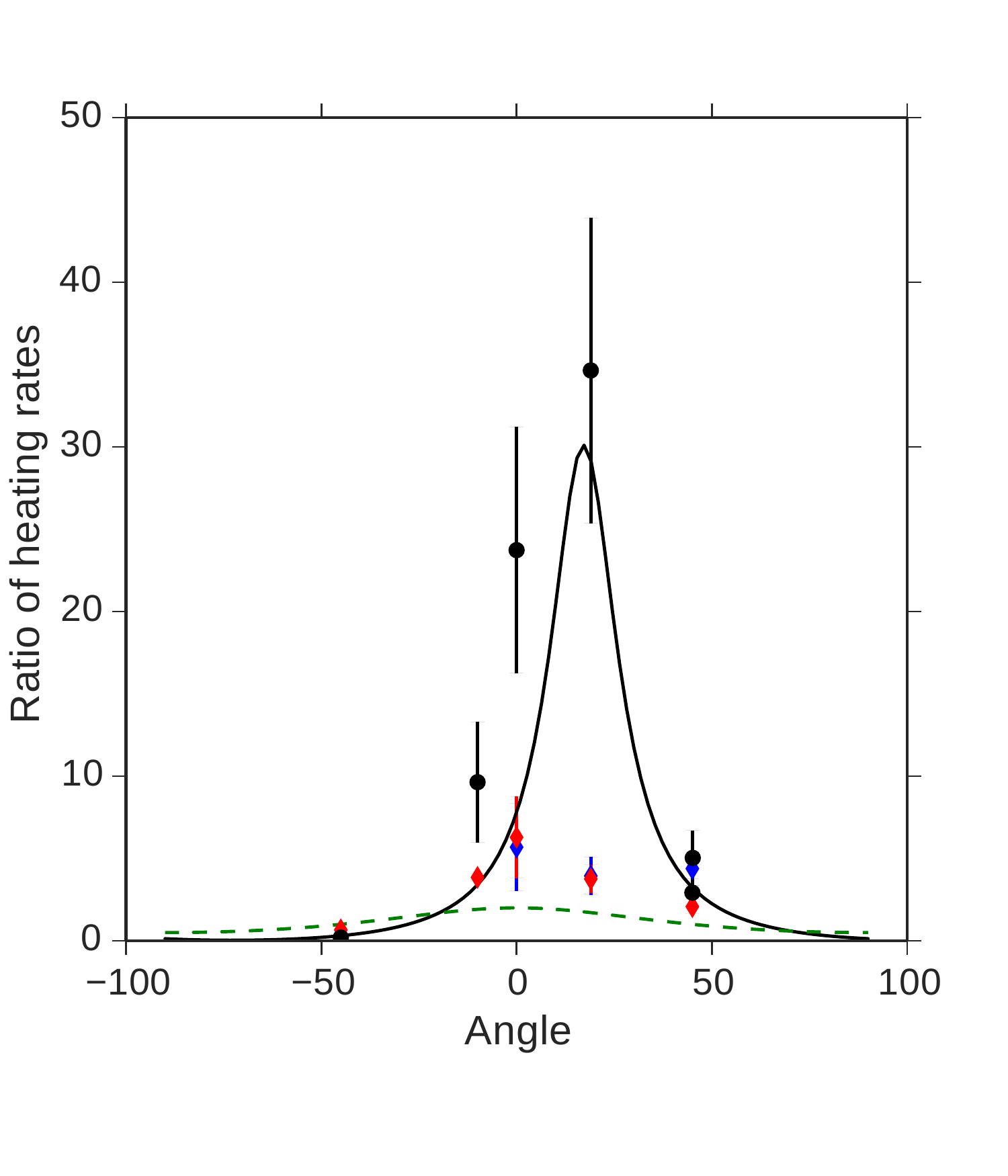}
  \caption{(color online) Measured ratio of the heating rate in the
    two radial modes. Red and blue diamonds are measurements without
    additional noise taken on different days. Black crosses are
    measurements with additional noise on the central electrode. The
    green dashed line corresponds to surface noise whereas the black
    solid line corresponds to the technical noise model. }
  \label{fig:ratio_exp}
\end{figure}

We measured the heating rates in both modes while keeping the trap
frequencies constant at $2.6\pm0.1$~ MHz. Figure \ref{fig:ratio_exp}
shows the heating rate as a function of the normal mode angle. For all
but the angle $\phi_g$ the axes are rotated by controlling the static
multipole confinement of all electrodes. For measuring at trap
orientation with angle $\phi_g$, a static bias voltage is applied to
the RF electrodes.

Applying this bias voltage on the RF electrodes allows for the
most reliable trap rotation and thus we will only use this method for
quantitative analysis of the noise polarization. We find a ratio of
heating rates in the two modes of $R=\rval$ which is small compared to
the ratio predicted by the voltage independent noise model of $R
\approx 30.1$. From this we can exclude the voltage independent
technical noise model as the dominant noise source in our setup.

In order to exclude the voltage dependent noise model, we measure the
heating rate for the $\updownarrow$ mode for two different sets of
voltages providing an axial confinement of approximately 1~MHz (for
set {\it i}) and 707~kHz (for set {\it ii}) while keeping the radial
trap frequencies constant. The voltages of the sets differ by a factor
of two and assuming the voltage noise to be proportional to the
voltage, one would expect the heating rates to differ by a factor of
four as the heating rate scales with the power spectral density of the
noise. We measure a heating rate of $\dot{n}=0.69(6)$~quanta/ms for
set {\it i} and $\dot{n}=0.52(3)$~quanta/ms for set {\it ii}, yielding
a factor of 1.3(1) between the two heating rates. With this result we
can exclude being dominated by noise that scales linearly with the
applied voltage, as the model predicts a change in heating rate of a
factor of four. A weaker scaling cannot be excluded completely but
inferring a scaling factor would give no meaningful results due to
large statistical uncertainties. 

We further test the method by adding voltage noise to only the central
electrode with a white noise generator. This should lead to an
increase of the heating rate in the mode parallel to the maximum noise
direction, whereas the perpendicular mode should not be affected. The
experiments demonstrate this effect: The heating rate in the
perpendicular mode without adding noise is $\dot{n}_\leftrightarrow =
0.12(3)$~quanta/ms and with added noise $\dot{n}_\leftrightarrow =
0.15(3)$~quanta/ms. In contrast, the measured heating rates in the
perpendicular mode are $\dot{n}_\updownarrow = 1.3(5)$~quanta/ms and
with added noise $\dot{n}_\updownarrow=5.5(1)$~quanta/ms. This
indicates that we can align the trap axes with the electric field
generated by the center electrode (at angle $\phi_g$) with adequate
precision.

\section{Estimating surface noise in the presence of technical noise}

Assuming a surface noise model featuring a ratio of $R_{\rm surf}=2$,
we can estimate the magnitude of surface noise even in the presence of
technical noise yielding a ratio $R_{\textrm techn}$. It is convenient
to perform the measurement at angle $\phi_g$ as this angle can be set
with highest precision. Assuming that surface and technical noise are
not correlated, the noise power spectral density is additive (the
fields add in squares): $S_{\rm tot} =  (S_{\rm surf} +
S_{\rm techn})$ with $S_{\rm techn}$ originating from voltage
independent technical noise. The ratio of the heating rates between
both axes is measured and thus it is possible to estimate the
magnitude of the surface noise as
\[S_{{\rm surf},\leftrightarrow} = S_{{\rm tot},\leftrightarrow} \,
\frac{R_{\rm tot}-R_{\rm techn}}{R_{{\rm surf},\phi}-R_{\rm techn}} \;
.\] For the measured ratio $R_{\rm tot}=\rval$ and the expected ratio
for patch potentials $R_{{\rm surf},\phi}=R_{\rm surf} \cos{^2\phi} =
2 \, \cos{^2\phi}$, this leads to $S_{{\rm surf},\leftrightarrow} =
1.8(2) \times 10^{-12} {\rm(V/m)^2/Hz}$. One needs to keep in mind that
this noise amplitude is measured at the angle $\phi_g$. The surface
noise magnitude parallel to the trap surface (along the x-axis) is
then
\begin{multline}
S_{{\rm surf},x} = \frac{
  S_{{\rm surf},\leftrightarrow}}{R_{\rm surf} \sin(\phi)^2 + \cos(\phi)^2} = \\
 1.7(2) \times 10^{-12} ({\rm V/m})^2{/Hz} \; .
\end{multline}

\section{Summary and conclusions}

We studied the noise polarization and found a factor of \rval of the
noise between the normal modes at an angle of $\phi_g$.  If we assume
a technical noise model where all electrodes show the same noise
amplitude, we expect a factor of 30. From this we can exclude that the
electric field noise parallel to the surface is dominated by such a
noise source. An alternative noise model, where the noise depends on
the applied voltage on the electrode, can lead to polarization similar
to the one that is expected from surface noise. This noise model can
be ruled out by comparing the noise magnitude for different applied
voltages.

While it is certainly possible to construct technical noise models
which explain our results by carefully choosing the amplitudes and
correlations of the various voltage supplies, those models seem rather
contrived. Assuming a simple and realistic technical noise model and a
surface noise caused by either surface dipoles or fluctuating patch
potentials, we can disentangle the contributions from technical noise
and surface noise. From this we can conclude that, in our setup,
technical noise is irrelevant to the field noise parallel to the trap
surface, while its contribution in the vertical direction is
comparable to surface noise. Using this method it will be possible to
compare heating rates from different traps, allowing a meta-analysis
of different experiments. We anticipate that this technique will be a
useful tool towards solving the mystery behind the unexpected noise
small ion traps suffer from \cite{Brownnutt2014}.

\section*{Acknowledgements}

The authors thank M. Brownnut, M. Kumph and P. Rabl for helpful
discussions. P.S. is supported by the Austrian Science Foundation
(FWF) Erwin Schr\"odinger Stipendium 3600-N27. This work has been
supported by AFOSR through the ARO grant FA9550-11-1-0318.  This
research was partially funded by the Office of the Director of
National Intelligence (ODNI), Intelligence Advanced Research Projects
Activity (IARPA), through the Army Research Office grant
W911NF-10-1-0284. All statements of fact, opinion or conclusions
contained herein are those of the authors and should not be construed
as representing the official views or policies of IARPA, the ODNI, or
the U.S. Government.

\bibliographystyle{apsrev4-1}
\bibliography{noise_polarization}

\end{document}